\documentstyle[epsfig,12pt]{article}

\begin{document}
                Submitted to "Foundation of Physics"
\bigskip
\begin{center}

{\bf
\title "Astrophysical Effects Related to Gravity-Induced Electric Polarization of Matter.}

\bigskip
\author "B.V.Vasiliev
\bigskip

Institute in Physical and Technical Problems,Dubna,Russia,141980
\bigskip

$e-mail: vasiliev@main1.jinr.ru$
\end{center}

\bigskip
\begin{abstract}
The calculations in Thomas-Fermi approximation show that in a
gravitational field each cell of ultra dense matter inside
celestial bodies obtains a very small positive electric charge. A
celestial body is electrically neutral as a whole, because the
negative electric charge exists at its surface. The positive
volume charge is very small, on the order of magnitude it equals
to $10^{-18}e$ per atom only. But it is sufficient to explain the
occurrence of magnetic fields of the celestial bodies and the
existence of a discrete spectrum of steady-state values of masses
of planets, stars, and pulsars.
\end{abstract}


\clearpage

\section{Introduction}
According to the conventional point of view, gravity does not
induce any electric polarization in the interior of celestial
bodies and electric forces are never considered in the balance of
matter of celestial bodies. Moreover, it is generally assumed that
the electric interaction plays practically no role in
astrophysics. It is a consequence of the comprehension that the
appreciable electric polarization cannot arise in metals and other
nonsegneto- and nonpiro-electric materials. It is entirely correct
for all substances under action of small pressure. But, thus, one
can disregard the fact that ultrahigh pressure transmutes all
substances into plasma state and radically changes the properties
of substance. In ultradense plasma, there is a different additional
mechanism of the gravity-induced electric polarization.

In a large celestial body, consisting of ultradense plasma, this
gravity-induced electric polarization (GIEP) can be rather great
and can play a determining role in the formation of a number of
features of the structure of a celestial body and its properties.

First of all, it concerns the following three problems, the
statement and the solution of which change drastically:

- the distribution of pressure and density of matter inside a
celestial body;

- the generation of a magnetic field by celestial bodies;

- the formation of a spectrum of steady-state values of masses of
celestial bodies.

As a consequence, these features of the structure can influence
the evolution of stars.

\section{The gravity-induced electric polarization in conducting matter}
The action of gravity on metals has often been a topic of
discussion before \cite{1}-\cite{6}. The basic result of these
researches is reduced to the statement that inside
a metal gravity induces  an electric field with an intensity

\begin{equation}
\bf{E}\simeq\frac{m_{i}\bf{g}}{e},\label{1010}
\end{equation}

where $m_{i}$ is the mass of an ion,

$\bf{g}$ is gravity acceleration,

$e$ is the electron charge.

This field is so small that it is not possible to measure it
experimentally. It is a direct consequence of the presence of an
ion lattice in a metal. This lattice is deformed by gravity and
then the electron gas adapts its density to this deformation. The
resulting field becomes very small.

Under superhigh pressure, all substances transform into ultradense
matter usually named nuclear-electron plasma \cite {7}. It occurs
when external pressure enhances the density of matter several
times \cite{7,8}. Such values of pressure exist inside celestial
bodies.

In nuclear-electron plasma the electrons form the degenerated
Fermi gas. At the same time, the positively charged ions form
inside plasma a dense packing lattice \cite{9},\cite{10}. As
usually accepted, this lattice may be replaced by a lattice of
spherical cells of the same volume. The radius $r_{s}$ of such a
spherical cell in plasma of the mass density $\gamma$ is given by

\begin{equation}
\frac{4\pi}{3}r_{s}^{3}=\biggl(\frac{\gamma}{m_{i}}\biggr)^{-1}=
\frac{Z}{n},\label{1020}
\end{equation}

where Z is the charge of the nucleus, $m_{i}=Am_{p}$ is the mass
of the nucleus, A is the atomic number of the nucleus, $m_{p}$ is
the mass of a proton, and n is the electron number density

\begin{equation}
n=\frac{3Z}{4\pi{r_{s}^{3}}}.\label{1030}
\end{equation}

The equilibrium condition in matter is described by the constancy
of its electrochemical potential \cite{7}. In plasma, the direct
interaction between nuclei is absent, therefore the equilibrium in
a nuclear subsystem of plasma (at $T=0$) looks like

\begin {equation}
\mu_{i}=m_{i}\psi+Ze\varphi=const.\label{1040}
\end {equation}

Here $\varphi$ is the potential of an electric field and $\psi$ is
the potential of a gravitational field.

The direct action of gravitation on electrons can be neglected.
Therefore, the equilibrium condition in the electron gas is

\begin {equation}
\mu_{e}=\frac{p_{F}^{2}}{2m_{e}}-(e-\delta{q})\varphi=const,\label{1050}
\end {equation}

where $m_{e}$ is the mass of an electron and $p _ {F} $  is the
Fermi momentum.

By introducing the charge $\delta {q}$, we take into account that
the charge of the electron cloud inside a cell can differ from
$Ze$. A small number of electrons can stay at the surface of a
plasma body where the electric potential is absent. It results
that the charge in a cell, subjected to the action of the electric
potential, is effectively decreased on a small value $\delta {q}$.
If the radius of a star $R_{0}$ is approximately ${10^{10}cm}$,
one can expect that this mechanism gives on the order of magnitude
$\frac{\delta {q}}{e}\simeq\frac{r_{s}}{R_{0}}\simeq{10^{-18}}$.

The electric polarization in plasma is a result of changing in
density of both nuclear and electron gas subsystems. The
electrostatic potential of the arising field is determined by the
Gauss' law

\begin{equation}
\nabla^{2}\varphi=\frac{1}{r^{2}}\frac{d}{dr}\biggl[r^{2}\frac{d}
{dr}\varphi\biggr]= -4\pi\biggl[Ze\delta(r)-en\biggr],\label{1060}
\end{equation}

where the position of nuclei is described by the function
$\delta(r)$.

According to the Thomas - Fermi method, $n$ is approximated by

\begin{equation}
n=\frac{8\pi}{3h^{3}}p^{3}_{F}.\label{1070}
\end{equation}

With this substitution, Eq.({\ref{1060}}) is converted into a
nonlinear differential equation for $\varphi$, which for $r>0$ is
given by

\begin{equation}
\frac{1}{r^{2}}\frac{d}{dr}\left(r^{2}\frac{d}{dr}\varphi(r)\right)=
4\pi\left[\frac{8\pi}{3h^{3}}\right] \left[2m_{e}(\mu_{e}+(e-
\delta{q})\varphi)\right]^{3/2}.\label{1080}
\end{equation}

It can be simplified by introducing the following variables
\cite{11}:

\begin{equation}
\mu_{e}+(e-\delta{q})\varphi=Ze^{2}{\frac{u}{r}}\label{1090}
\end{equation}

and $r=ax$,

where

$a=\{\frac{9\pi^{2}}{128Z}\}^{1/3}a_{0}$

with $ a_{0}=\frac{\hbar^{2}}{m_{e}e^{2}}=$ Bohr radius.

With the account of Eq.({\ref{1040}})

\begin{equation}
Ze^{2}{\frac{u}{r}}= const
-\frac{m_{i}\psi}{Z}-\delta{q}\varphi.\label{1092}
\end{equation}

Then Eq.({\ref{1080}}) gives

\begin{equation}
\frac{d^{2}u}{dx^{2}}=\frac{u^{3/2}}{x^{1/2}}.\label{1100}
\end{equation}

In terms of u and x, the electron density within a cell is given
by \cite{11}

\begin{equation}
n_{TF}=\frac{8\pi}{3h^{3}}p^{3}_{F}=
\frac{32Z^{2}}{9\pi^{3}a^{3}_{0}}
\biggl(\frac{u}{x}\biggr)^{3/2}.\label{1110}
\end{equation}

Under the influence of gravity, the charge of the electron gas in a
cell becomes equal to

\begin{equation}
Q_{e}=4\pi{e}\int^{r_{s}}_{0}n(r)r^{2}dr=\frac{8\pi{e}}{3h^{3}}
\biggl[2m_{e}\frac{Ze^{2}}{a}\biggr]^{3/2}4\pi{a}^{3}
\int^{x_{s}}_{0}x^{2}dx\biggl[\frac{u}{x}\biggr]^{3/2}.\label{1140}
\end{equation}

Using Eq.({\ref{1100}}), we obtain

\begin{equation}
Q_{e}=Ze\int^{x_{s}}_{0}xdx\frac{d^{2}u}{dx^{2}}=
Ze\int^{x_{s}}_{0}dx\frac{d}{dx}\biggl[x\frac{du}{dx}-u\biggr]=
Ze\biggl[x_{s}\frac{du}{dx}\bigg|_{x_{s}}-u(x_{s})+u(0)\biggr].
\label{1150}
\end{equation}

At $ r\rightarrow0 $ the electric potential is due to the nucleus
alone $ \varphi{(r)} \rightarrow\frac {Ze} {r} $. It means that $
u(0)\rightarrow1 $ and each cell of plasma obtains a small charge

\begin{equation}
\delta{q}=Ze\biggl[{x_{s}}\frac{du}{dx}\bigg|_{x_{s}}-u(x_{s})\biggr]=
Ze{x_{s}}^2\biggl[\frac{d}{dx}\biggl(\frac{u}{x}\biggr)\biggr]_{x_{s}}.
\label{1160}
\end{equation}

For a cell placed in the point $\bf{R}$ inside a star

\begin{equation}
\delta{q}=Zer_{s}^2\biggl[\frac{d}{d\bf{R}}
\biggl(\frac{u}{r}\biggr)\biggr]\biggl[\frac{d\bf{R}}{dr_{s}}\biggr].
\label{1170}
\end{equation}

Considering that the gravity acceleration
$\bf{g}=-\frac{d\psi}{d\bf{R}}$ and the electric field intensity
$\bf{E}=-\frac{d\varphi}{d\bf{R}}$

\begin{equation}
\frac{dr_{s}}{d\bf{R}}=\frac{r_{s}^2}{e}\biggl[\frac{\frac{m_{i}}{Z}\bf{g}
+\delta{q}\bf{E}}{\delta{q}}\biggr].\label{1180}
\end{equation}

This equation has the following solution

\begin{equation}
\frac{dr_{s}}{d\bf{R}}=0\label{1190}
\end{equation}

and

\begin{equation}
\frac{m_{i}}{Z}\bf{g}+\delta{q}\bf{E}=0.\label{1200}
\end{equation}

In plasma, the equilibrium value of the electric field on nuclei
according to Eq.({\ref{1040}}) is determined by Eq.({\ref{1010}})
as well as in a metal. But there is one more additional effect in
plasma. Simultaneously with the supporting of nuclei in
equilibrium, each cell obtains an extremely small positive
electric charge.

As $div{\bf{g}}=-4\pi{G}{n}m_{i}$ and
$div{\bf{E}}=4\pi{n}\delta{q}$, the gravity-induced electric
charge in a cell

\begin{equation}
\delta{q}=\sqrt{G}\frac{m_{i}}{Z}\simeq{10^{-18}e},\label{1210}
\end{equation}

where $G$ is the gravity constant.

However, because the sizes of bodies may be very large, the
electric field intensity may be very large as well

\begin{equation}
\bf{E}=\frac{\bf{g}}{\sqrt{G}}.\label{2050}
\end{equation}

In accordance with Eqs.({\ref{1190}},{\ref{1200}}), the action of
gravity on matter is compensated by the electric force and the
gradient of pressure is absent.

Thus, a celestial body is electrically neutral as a whole, because
the positive volume charge is concentrated inside the charged core
and the negative electric charge exists on its surface and so one
can infer gravity-induced electric polarization  of a body.

\section{Pressure distribution inside a celestial body.}

As at the surface of a celestial body pressure is absent, near
this surface there is always a stratum where plasma and
polarization are absent. For the large stars, the size of this stratum
is insignificant. But for a small planet it can comprise a
substantial part of a planet, and thus, only a small relatively
internal region will be polarized. At the surface of this core,
the electric field intensity falls to zero. The jump in the
electric field intensity is accompanied on the surface of the core
by the pressure jump $\Delta p(R_{N})$(\cite{12}-\cite{13}). The
important astrophysical consequence of the GIEP effect is the
redistribution of the matter density inside a celestial body. In a
celestial body, consisting of matter with an atomic structure,
density and pressure grow monotonously with depth. In a celestial
body, consisting of electron-nuclear plasma, the GIEP effect results
in the fact that the pressure gradient inside the polarized core
is absent and the matter density is constant. Pressure affecting the
matter inside this body is equal to the pressure jump at the
surface of the core

\begin{equation}
p=\Delta p(R_{N})=\frac{E(R_{N})^{2}}{8\pi }=\frac{2\pi
}{9}G\gamma ^{2}R_{N}^{2},  \label{3010}
\end{equation}

where $\gamma$ is the matter density in the core and R$_{N}$ is
the radius of the core.

One can say that this pressure jump is due to the existence of the
polarization jump or, which is the same, the existence of the
bonded surface charge, which is formed by electron pushed out from
the core and which makes the total charge of the celestial body
equal to zero.

\section{Earth's structure.}

It is important, that the GIEP effect gives the possibility to
construct the intrinsically self-consistent theory of the Earth
\cite{13}. Although it is rather a solution of a geophysical
problem than an astrophysical effect.

Earlier models of the Earth assumed the existence of the
monotonous dependence of pressure inside the planet. The division
of the Earth into the core and the mantle was explained by the
fact that at the creation of the Earth, on its share a certain
amount of iron (and other heavy metals) and also a necessary
amount of stone were given out. The core consists of metals and
the mantle consists of stone. In these models, it was necessary to
fit the parameters to get the densities of core and mantle and
their sizes. It is not necessary to introduce any free parameters
into the Earth theory based on the GIEP effect. Assuming that the
Earth consists of homogeneous matter, the division on core and
mantle is explained by the existence of the pressure jump on the
surface of the core Eq.({\ref{3010}}). The basic results of this
theory are reduced to the calculation of the following five values:

a) the radius of the Earth's core;

b) the density of core matter;

c) the density jump on the core-mantle boundary;

d) the mass related to one electron of the Fermi gas in the core;

e) the electric polarization of the core.

To express it in appropriate equations, one should substitute the
following four parameters (the gravitational constant $G$ is
known):

\begin{figure}
\begin{center}
\includegraphics{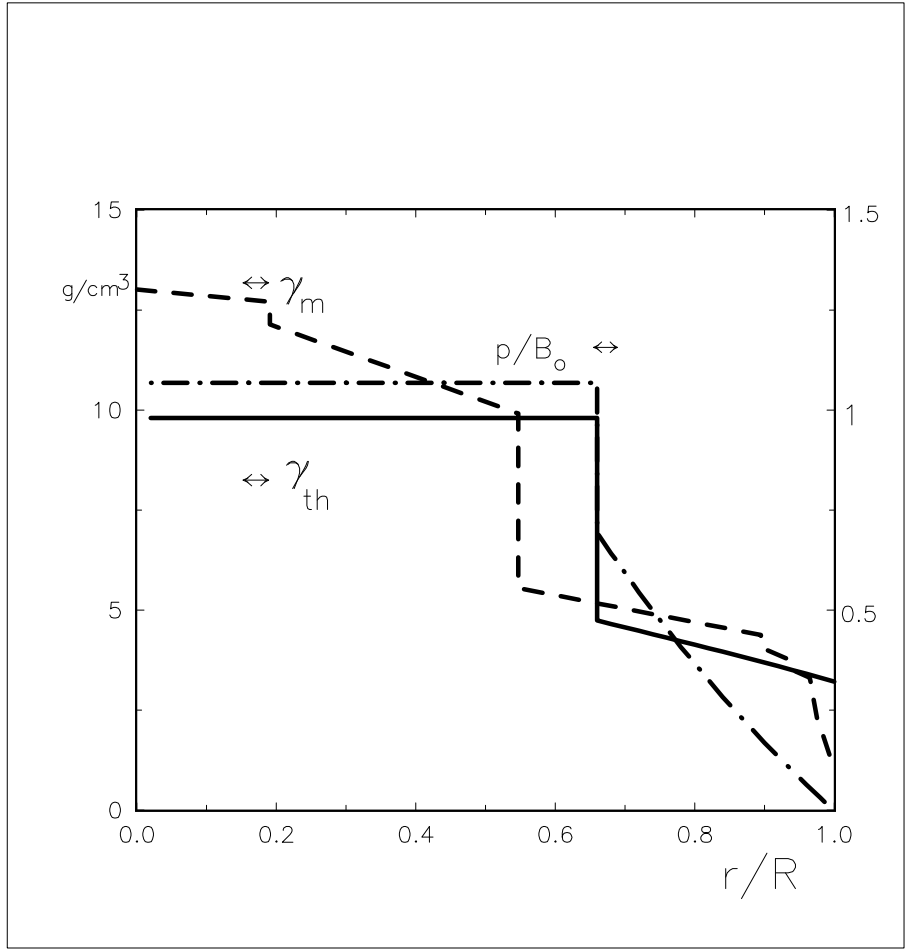}
\caption{The radial dependence of pressure and the matter density
inside the Earth. The solid line is the calculated dependence of
the matter density; the dashed line is the density of the Earth
obtained by measuring the propagation velocity of seismic waves.
The dash-dotted line is the calculated dependence of pressure
inside the Earth over bulk module B=$1.3\cdot 10^{12}$
dyn/cm$^{2}$.} \label{fig1}
\end{center}
\end{figure}

a) the mass of the Earth;

b) the radius of the Earth;

c) the matter density on the surface of the Earth;

d) the bulk module of matter at the surface of the Earth.

Thus, other parameters can be obtained, for example, the pressure
distribution inside the Earth. The basic results of this theory are shown in
Fig.1.

In addition, from the obtained data it is possible to calculate the
angular momentum of the Earth. This calculation gives the value of
$0.339MR^2$. It is in agreement with the measured value of
$0.331MR^2$ within several percent of the accuracy.

It is possible to calculate the  magnetic moment of the Earth.

Apparently, using the appropriate data of other planets (the mass,
the size, and the properties of matter at the surface), it is
possible to construct models of these planets. It can be made, if
these planets have electrically polarized cores and corresponding
magnetic fields.

\section{The gyromagnetic ratio of a celestial body}

Another astrophysical consequence of the GIEP effect is coupled
by the rotation of celestial bodies about their axes. A celestial
body is electroneutral as a whole. The positive volume charge is
concentrated inside the core and the negative charge is located at
the surface of the core. When rotating, they move on different
radii. As a result, all celestial bodies, when the GIEP effect
is present, obtain magnetic moments

 \begin{equation}
 \mu=\frac{2}{15}\frac{4\pi}{3c}\rho\Omega{R_{N}^{5}}.\label{5010}
 \end{equation}

If the size of the body is sufficiently large, the core radius
$R_{N}$ does not differ significantly from its external radius R.
For this celestial body, the angular momentum of the core
coincides by the order of magnitude with the angular momentum of
the body as a whole

 \begin{equation}
 L=\frac{2}{5}M\Omega{R}^{2}\label{5020}
 \end{equation}

where $M=\frac{4\pi}{3}{\gamma{R^{3}}}$ is the mass of a celestial
body and $\Omega$ is the velocity of rotation.

\begin{figure}
\begin{center}
\includegraphics[5cm,14cm][17cm,2cm]{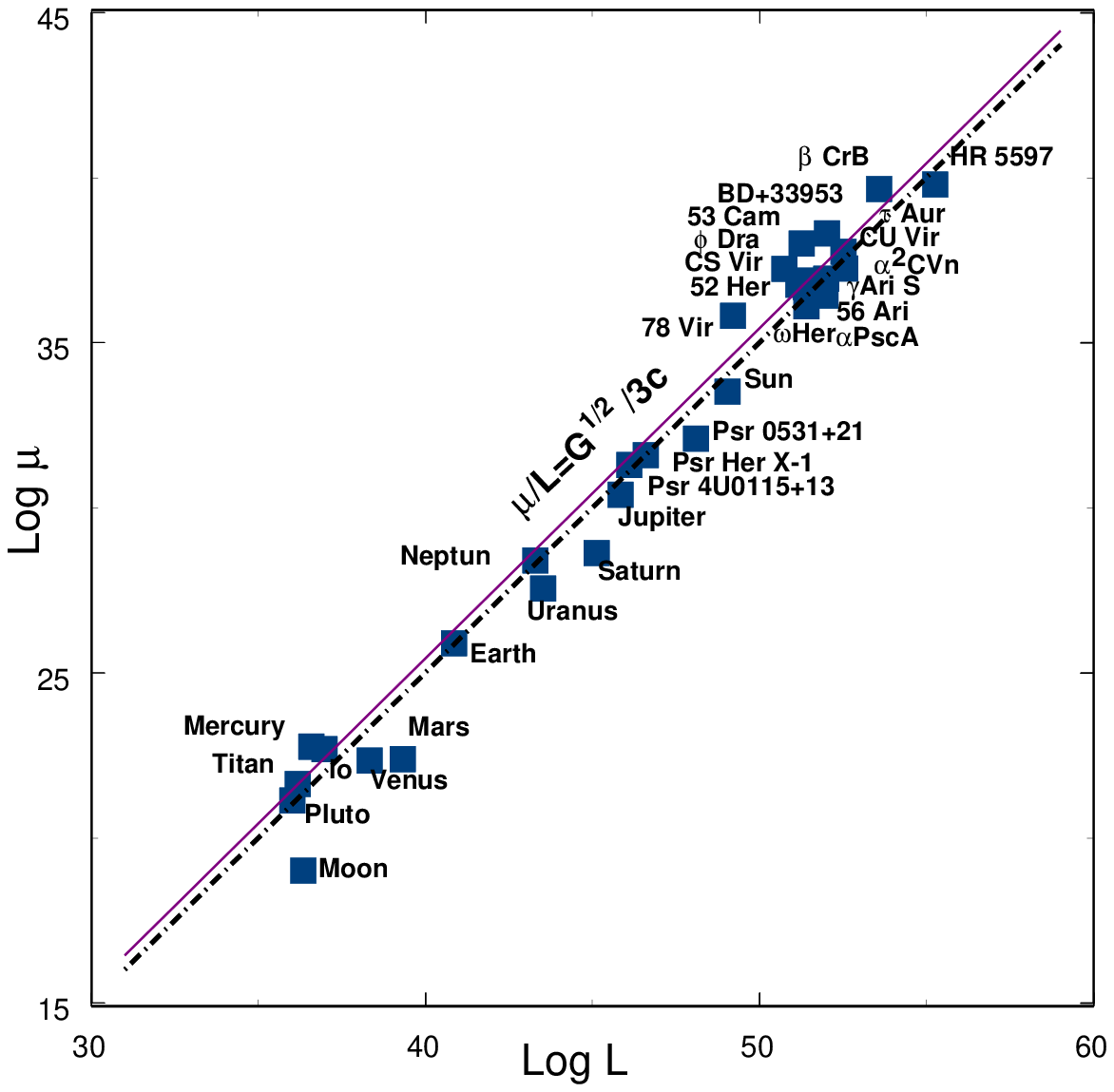}
\vspace{11cm} \caption{The observed values of the magnetic moments
of celestial bodies vs. their angular momenta. On the ordinate,
the logarithm of the magnetic moment over $Gs\cdot{cm^3}$ is
plotted; on the abscissa the logarithm of the angular momentum
over $erg\cdot{s}$ is shown. The solid line illustrates
Eq.({\ref{5030}}). The dash-dotted line is the fitting of the
observed values.} \label{fig2}
\end{center}
\end{figure}

\begin{figure}
\begin{center}
\includegraphics[5cm,14cm][17cm,2cm]{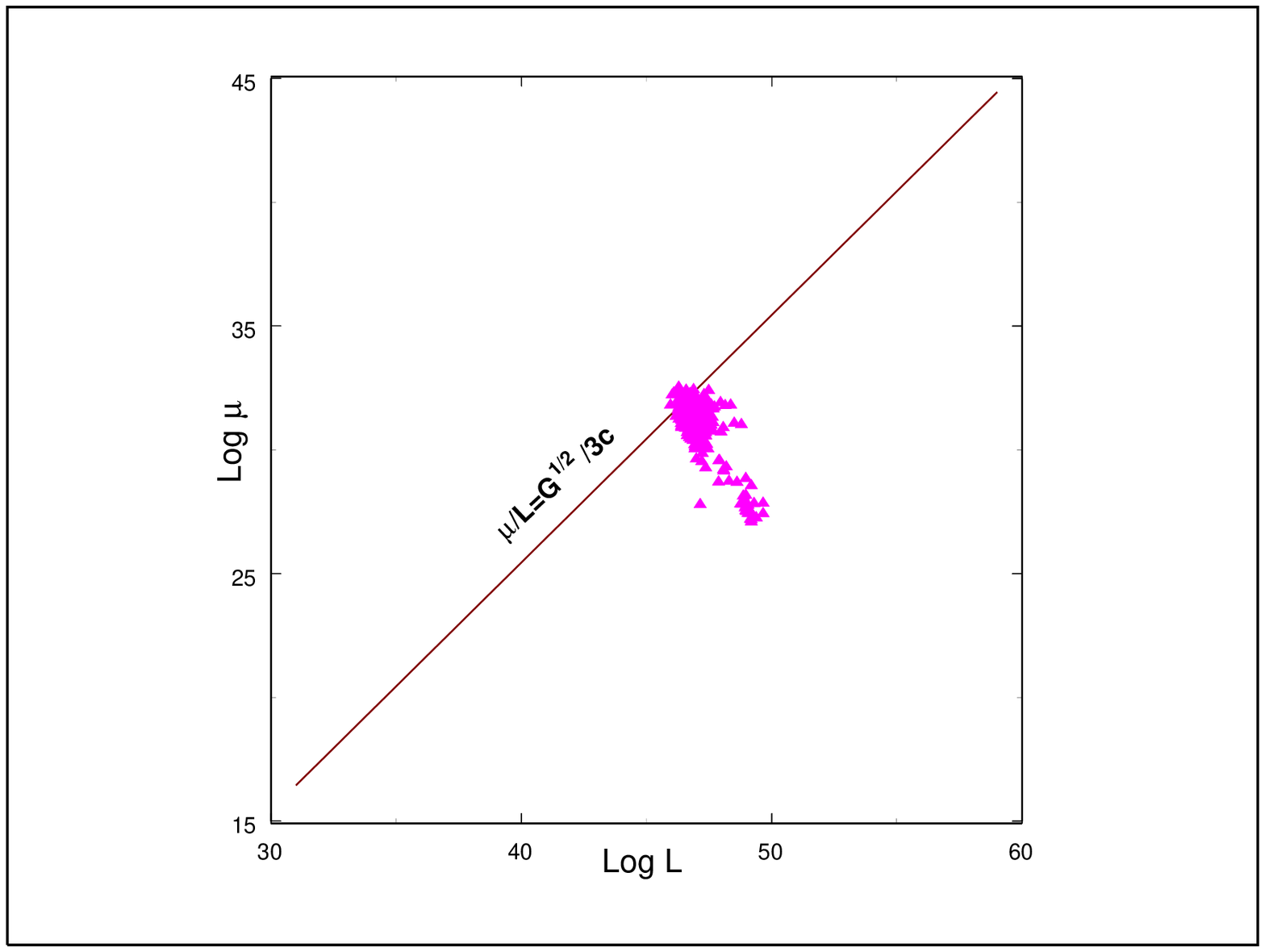}
\vspace{14cm} \caption{The estimated values of the magnetic
moments of pulsars \cite{19} vs. their angular momenta. Solid line
is Eq.({\ref{5030}}). The axes are as in Fig.2.} \label{fig3}
\end{center}
\end{figure}

 Finally, the gyromagnetic ratios for these bodies should be close to
 the universal value

 \begin{equation}
 \frac{\mu }{L}=\frac{G^{1/2}}{3c}.\label{5030}
 \end{equation}

The values of $\mu $(L) for all celestial bodies (for which they
are known today) are shown in Fig.2. The data for planets are
taken from \cite{14}, the data for stars are taken from \cite{15},
and for pulsars - from \cite{16}.

As can be seen from the figure with the logarithmic accuracy, all
celestial bodies - stars, planets, and pulsars - really have the
gyromagnetic ratio close to the universal
value $\frac{G^{1/2}}{3c}$. Only the data for the Moon fall out,
because its size and inner pressure are too small to create an
electrically polarized core. The estimation of the  magnetic moment of
the Earth within the frame of the theory mentioned above \cite{13}
gives $\mu \simeq 4\cdot 10^{25}Gs\cdot cm^{3}$. It is almost
precisely one half from the observed value of $8.05\cdot
10^{25}Gs\cdot cm^{3}$. For some planets, the values of magnetic
moments are in a good agreement with Eq.({\ref{5030}}) but they
have an opposite sign. Apparently, it means that the hydrodynamic
mechanism also plays a certain role.

For the majority of pulsars, there are estimations of magnetic
fields \cite{19} obtained using a number of model assumptions
\cite{16}. It is impossible to consider these data as the data of
measurements, but nevertheless, they also agree in certain way
with Eq.({\ref{5030}}),(Fig.3)

\section{The masses of celestial bodies.}

The important astrophysical outcome of the GIEP effect is a
discrete distribution of masses of celestial bodies. This spectrum
is a result of the fact that electron-nuclear plasmas can exist in
various states.

The equation of state of matter subjected to high pressure is
usually described as a polytrope \cite{7}:

\begin{equation}
p=C\cdot \gamma ^{1+\frac{1}{k}},  \label{6010}
\end{equation}

where C is the dimensional constant,

k is the polytropy.

\subsection{Nonrelativistic electron-nuclear plasma.}

At relatively small pressure, substances are transmuted into
nonrelativistic electron-nuclear (or electron-ion) plasma. It is
peculiar to conditions existing inside cores of planets. According
to \cite{7}, the state equation of the nonrelativistic
electron-nuclear plasma (characterized by the polytropy k=3/2) is

\begin{equation}
p_{(3/2)}=\frac{\left( 3\pi ^{2}\right) ^{2/3}\hbar ^{2}\gamma ^{5/3}}{%
5m_{e}(\beta\cdot m_{p})^{5/3}},  \label{6020}
\end{equation}

where $\beta\cdot{m_{p}}$ is the mass of matter related to one
electron of the Fermi gas system and m$_{p}$ is the proton mass.

If the pressure inside a celestial body is formed by the GIEP
effect and is determined by Eq.({\ref{3010}}), than from
Eq.({\ref{6020}}) for the nonrelativistic Fermi gas of electrons,
we obtain the steady-state value of mass for a core of planet

\begin{equation}
M_{(3/2)}=C_{(3/2)}\cdot \left( \frac{\hbar
^{2}}{Gm_{e}m_{p}}\right) ^{3/2}\cdot \frac{\gamma ^{1/2}}{\beta
^{5/2}m_{p}},  \label{6030}
\end{equation}

where C$_{(3/2)}=\frac{54\pi }{5}\left( \frac{\pi }{10}\right)
^{1/2}\simeq 19.$

The dependence of Eq.({\ref{6030}}) is shown in Fig.4. Therefore,
any planet (even consisting from pure hydrogen) should have a mass
less than 10$^{31}g$ (if its density is approximately equal to $1
g/cm^3$).

\begin{figure}
\begin{center}
\includegraphics[5cm,14cm][17cm,2cm]{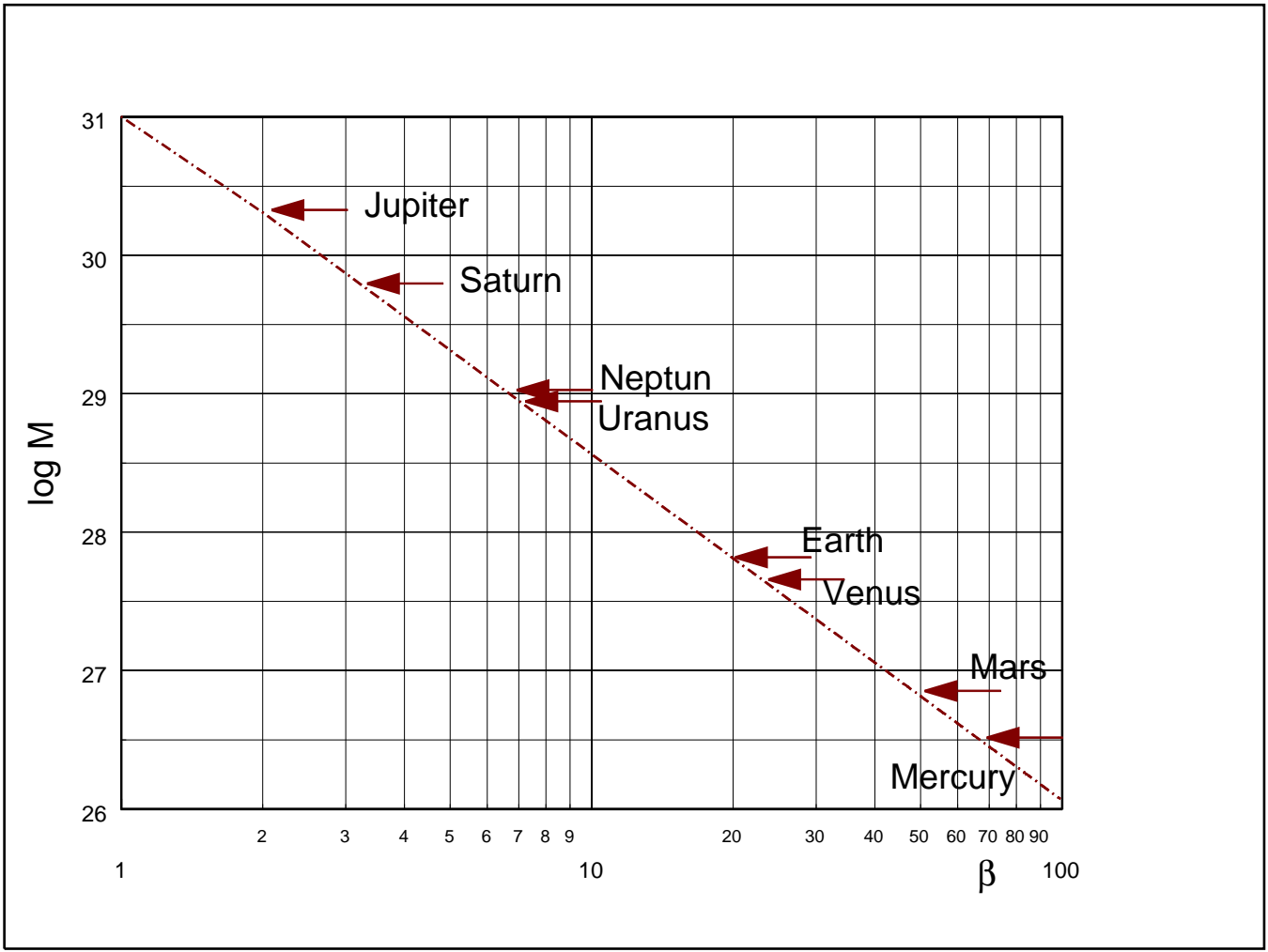}
\vspace{14cm}
\caption{The dependence of the core mass of planets
on $\beta$ (Eq.({\ref{6030}})) at $\gamma =1g/cm^{3}$. On the
ordinate, the logarithm of mass (over $1 g$) is plotted.}
\label{fig4}
\end{center}
\end{figure}

In Fig.4 the masses of the planets of the Solar system are marked.
The mass of the Jupiter is $1.9\cdot 10^{31}g$. It is close to the
specified limit. For the Jupiter Eq.({\ref{6030}}) gives $\beta
\simeq 2$. It is according to the data that the large planets have
the deuterium-helium composition. For other planets the mantle is
not small in comparison with their sizes. For this reason,
Eq.({\ref{6030}}) can give an excessive estimation for other
planets.

\subsection{Relativistic electron-nuclear plasma.}

When the pressure increases, the substances are transmuted into
relativistic electron-nuclear plasma (the polytropy k= 3). Its
state equation is \cite{7}

\begin{equation}
p_{(3)}=\frac{\left( 3\pi ^{2}\right) ^{1/3}\hbar c\gamma ^{4/3}}{%
4m_{p}^{4/3}\beta ^{4/3}}  \label{6040}
\end{equation}

If this plasma is originated by the GIEP effect, then the
steady-state value of mass of a star consisting of it, according
to Eqs.({\ref{3010}},{\ref{6040}}) is

\begin{equation}
M_{(3)}=C_{(3)}\cdot A_{\star}^{3/2}\cdot \frac{m_{p}}{\beta
^{2}}, \label{6050}
\end{equation}

where the dimensionless constants are

\begin{equation}
A_{\star}=\left( \frac{\hbar c}{Gm_{p}^{2}}\right) =1.54\cdot
10^{38} \label{6060}
\end{equation}

and C$_{(3)}=\left( 1.5^{5}\pi \right) ^{1/2}\simeq 4.88.$

Because of the electric neutrality, one proton should be related to
electron of the Fermi gas of plasma. The existence of one neutron
per proton is characteristic for a substance consisting of light
nuclei. The quantity of neutrons grows approximately to 1.8 per
proton for the  heavy nuclei substance. Therefore, it is necessary
to expect that inside stars $2<\beta<2.8$.

The masses of stars can be measured with a considerable accuracy,
if these stars compose a binary system. There are almost 200
double stars which masses are known with the required accuracy
\cite{17}. Among these stars there are giants, dwarfs, and stars of
the main sequence. Their average masses are described by the
equality

\begin{equation}
\langle M_{star}\rangle =\left( 1.36\pm 0.05\right) M_{\odot },
\label{6070}
\end{equation}

where $M_{\odot }$ is the mass of the Sun.

The center of this distribution (Fig.5) corresponds to
Eq.({\ref{6050}}) at $\beta \simeq 2.6$.

\begin{figure}
\begin{center}
\includegraphics[5cm,14cm][17cm,2cm]{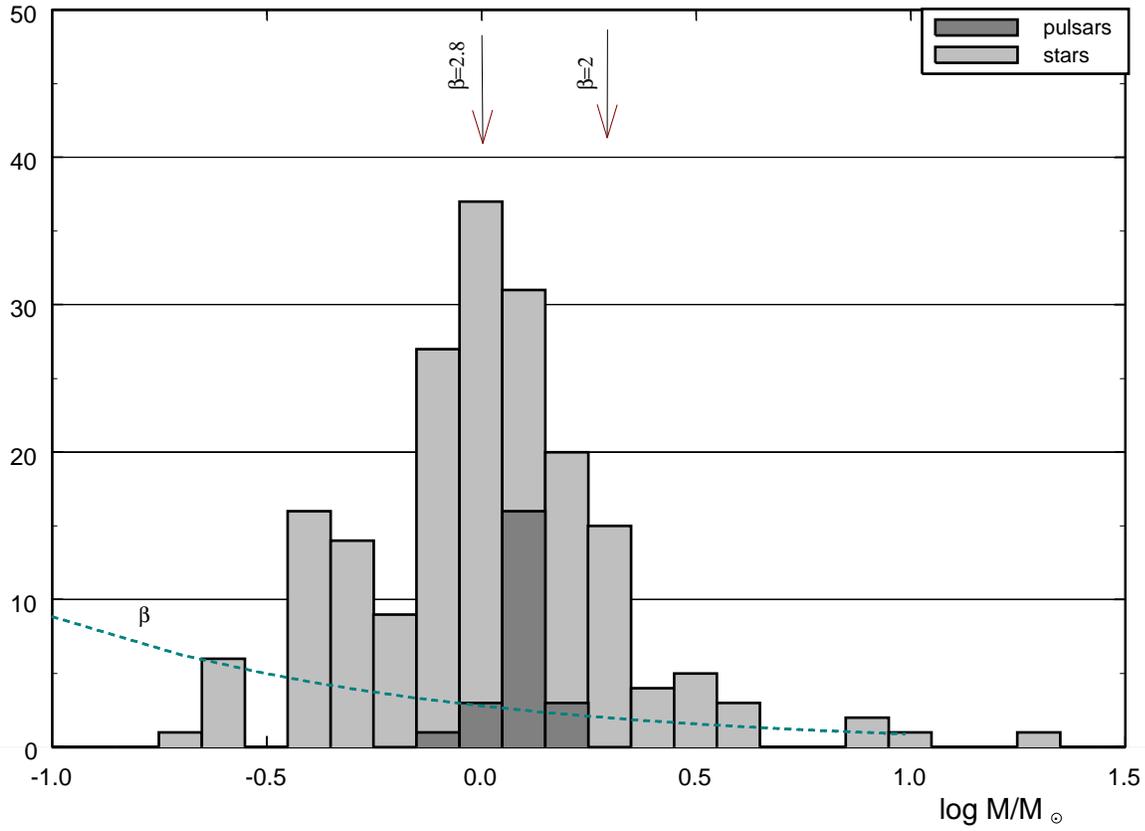}
\vspace{11cm}
\caption{Mass distributions of stars and pulsars
from the binary systems \cite{17}-\cite{18}. The curve shows
$\beta$ (Eq.({\ref{6050}})).} \label{fig5}
\end{center}
\end{figure}

\subsection{Ultrarelativistic electron-nuclear plasma.}

Further increase in pressure transmutes substances into
ultrarelativistic plasma. Then nuclear reactions of capture of
electrons by nuclei become favorable and the neutronization of
matter takes place. Equilibrium pressure of ultrarelativistic
plasma does not depend on its density. It is formally
characterized by the polytropy k=-1 and its state equation is
\cite{7}

\begin{equation}
p_{(-1)}=\frac{\Delta ^{4}}{12\pi ^{2}\left( \hbar c\right) ^{3}}.
\label{6080}
\end{equation}

Here $\Delta $ is the difference between the energy of the
initial nucleus and the energy of the daughter nucleus.

The equilibrium mass of a star, consisting of ultrarelativistic
plasma, according to Eqs.({\ref{3010}}),Eq.({\ref{6080}}) is

\begin{equation}
M_{(-1)}=C_{(-1)}\left( \frac{\Delta ^{6}}{\left( \hbar c\right)
^{9/2}G^{3/2}\gamma ^{2}}\right) ,  \label{6090}
\end{equation}

where C$_{(-1)}=\frac{1}{4\pi ^{3}}\left( \frac{3}{2\pi }%
\right) ^{1/2}\simeq 6\cdot 10^{-3}.$

According to the astrophysical data a neutronization of matter
takes place at the density $\gamma\approx{10^{7}}\frac{g}{cm^3}$.
Thus, Eq.({\ref{6090}}) gives

\begin{equation}
M_{(-1)}\approx{8\cdot{10^{32}}}g\approx{0.4}M_{\odot}.\label{6100}
\end{equation}

Certainly this result is the rough estimation on the order of
magnitude only, but it is in a satisfactory agreement with
measurements of the astronomers related to masses of white dwarfs
from double systems.

\subsection{Nonrelativistic neutron matter.}

At higher pressure, the substance is transmuted into a
nonrelativistic neutron matter with a small impurity of protons
and electrons \cite{7}. The state equation of the nonrelativistic
neutron matter will coincide with Eq.({\ref{3010}}) with a
replacement of $m_{e}$ with $m_{p}$

\begin{equation}
p_{(3/2)}^{(n)}=\frac{\left( 3\pi ^{2}\right) ^{2/3}\hbar ^{2}\gamma ^{5/3}}{%
5m_{p}^{8/3}\beta ^{5/3}}.  \label{6110}
\end{equation}

Together with Eq.({\ref{3010}}), it gives the equilibrium mass of
the nonrelativistic neutron star

\begin{equation}
M_{(3/2)}^{(n)}=C_{(3/2)}\left( \frac{\hbar ^{2}}{G}\right) ^{3/2}\frac{%
\gamma ^{1/2}}{m_{p}^{4}\beta ^{5/2}}.  \label{6120}
\end{equation}

As the density $\gamma \simeq 4\cdot 10^{13}g/cm^{3}$ and
$\beta=2.6$

\begin{equation}
M_{(3/2)}^{(n)}\simeq{0.05}M_{\odot }.\label{6130}
\end{equation}

The astronomers have not found such neutron stars.

\subsection{Relativistic neutron matter.}

With further increase in pressure, the neutron Fermi gas becomes a
relativistic one. Its state equation completely coincides with the
state equation of the relativistic Fermi gas of electrons
Eq.({\ref{3010}})

\begin{equation}
p_{(3)}^{(n)}=\frac{\left( 3\pi ^{2}\right) ^{1/3}\hbar c\gamma ^{4/3}}{%
4m_{p}^{4/3}\beta ^{4/3}}.  \label{6140}
\end{equation}

As well as the masses of relativistic stars, the masses of
relativistic pulsars do not depend on their density and can be
directly expressed through world constants

\begin{equation}
M_{(3)}^{(n)}=C_{(3)}\cdot A_{\star}^{3/2}\cdot \frac{m_{p}}{\beta
^{2}} \label{6150}
\end{equation}

(at $\beta =1$ for the pure neutron Fermi gas).

As it is specified in \cite{7}, at this density of matter it
is necessary to take into account nuclear forces and the presence
inside nuclear matter except neutrons also of protons, $\pi
^{-}$mesons, and electrons. It can be made using $\beta $ as a
parameter of the correction.

The mass of the neutron star can be measured with a considerable
accuracy if it enters into a binary system. The astronomers have
found 16 radio-pulsars \cite{18} and 7 X-pulsars \cite{16} in
binary systems. They all are located in a very narrow mass
interval (Fig.5)

\begin{equation}
\langle M_{pulsar}\rangle =(1.38\pm 0.03)M_{\odot }. \label{6160}
\end{equation}

The center of this distribution corresponds to Eq.({\ref{6150}})
with the correction parameter $\beta \simeq 2.6$ just as for
relativistic stars. Thus, we come to the conclusion that
Eq.({\ref{6150}}) on the order of magnitude correctly describes
the results of astronomical observations.

\section{Conclusion.}

It is expedient  to underline the basic obtained results in
summary.

1. The developed theory defines a concept of the steady-state
values of masses of celestial bodies related to their equations of
state and gives the possibility to calculate these values.

2. It gives the new way for the determination of the substance
density distribution inside celestial bodies. According to early
models, it was supposed that density of a substance inside celestial
bodies grows more or less monotonically with depth and at the
centre of a star, the density has the greatest value and even a
black hole may exist there. According to the developed theory,
the density of a substance inside a star is constant.

3. It is interesting to emphasize that the "biography" of such a
star appears much poorer than in the Chandrasecar model.

There cannot exist a black hole inside a star, and it should not
collapse with a temperature decrease. All the considered effects are
based on an equilibrium of the Fermi system. Temperature does not
influence the parameters of relativistic plasma. Therefore, a star
with a mass close to the steady-state value (Eq.({\ref{6050}})) is
in a stable equilibrium not depending on temperature. The existing
stars should retain the stability at any (even at zero)
temperature. Therefore, a collapse of the already existing stars
apparently is impossible. The instability of a star can arise with
burning out of light nuclei - deuterium and helium - and with a
related increasing of $\beta$. This growth leads to the reduction
of a steady-state value of mass (Eq.({\ref{6050}})) and, probably,
to the distraction of the stars wiht masses more than the
steady-state value.

4. The developed theory determines the simple and essential
mechanism of generation of the magnetic field by celestial bodies.
All early models tried to solve the basic problem - to calculate
the magnetic field of a celestial body. Such a statement of the basic
problem of planetary and stellar magnetism is unacceptable at present.
Space flights and a development of astronomy discovered a remarkable
and earlier unknown fact: the magnetic moments of all celestial bodies
are proportional to their angular momenta and the proportionality
coefficient is determined by the ratio of world constants. The
explanation of this phenomenon is the basic problem of planetary
and stars magnetism nowadays. Early models cannot explain this
phenomenon. The developed theory used for this explanation a
standard mechanism.

5. It is possible to consider that now the predicted steady-state
values of masses of celestial bodies and the predicted values of
their magnetic moments are in a satisfactory agreement with the
data of observations. But it is tempting to obtain these
data to closer limit of accuracy. Two arrows in the upper part of Fig.5 mark
masses of stars consisting of extremely light and heavy nuclei.
These values are obtained from Eq.({\ref{6050}}) without the use
of any fitting parameters. In agreement wiht the developed theory,
if stars have a "usual" chemical composition, there must be no
stars outside of this interval (or these stars should be
unstable). The histogram on Fig.5 is somewhat wider. It is
interesting to understand, whether there is a principal deviation
from the developed theory or it is a result of measuring errors.
First of all, it requires a more careful and precise measurement
of masses of binary stars.


\end{document}